# NETWORK INTERVENTIONS: APPLYING NETWORK SCIENCE FOR PRAGMATIC ACTION IN PUBLIC ADMINISTRATION AND POLICY

**Michael D. Siciliano***
University of Illinois Chicago, USA

**Travis A. Whetsell**
Georgia Institute of Technology, USA

**Abstract**: Public management and policy scholars have engaged in extensive development of theory and empirical study of networks and collaborative systems of governance. This scholarship has focused on understanding the mechanisms of network formation and the implications of network properties on individual and collective outcomes. Despite rich descriptive work and inferential analyses, little work has attempted to intervene in these systems. In this article, we develop the foundation for a new body of research in our field focused on network interventions. Network interventions are defined as the purposeful use of network data to identify strategies for accelerating behavior change, improving performance, and producing desirable outcomes (Valente, 2012). We extend network intervention strategies from the field of public health to public sector inter-organizational and governance networks. Public sector actors have an interest in network interventions based on the fundamental pursuit of efficiency, effectiveness, and equity. Network interventions can increase the uptake of an organizational change among employees, improve the performance of a governance system, or promote the spread of a successful policy across jurisdictions. We provide scholars and practitioners with a useful way to conceptualize where, why, and how network interventions might be deployed in the pursuit of public value.

## INTRODUCTION

Over the past few decades, a significant body of literature using network theory and analysis has emerged in public administration and policy (Hu, Medina, Siciliano, & Wang, 2022; Kapucu, Hu, & Khosa, 2017; Koliba, Meek, & Zia, 2011). This scholarship can be categorized into two types: (i) *network formation*, where network measures are the dependent variable and (ii) *network effect*, where network measures are the independent variable (Borgatti, Everett, & Johnson, 2013; Borgatti & Ofem, 2010). Research, where the network measures serve as the dependent variable, emphasizes the tie formation decisions of the actors (organizations, departments, individuals) in a social system (Nisar & Maroulis, 2017; Siciliano, Wang, & Medina, 2021; Whetsell, Kroll, & DeHart-Davis, 2021). Such studies often ask which actor attributes are most predictive of their position in a network or why some networks are more or less centralized than others. Thus, network formation studies treat the network's structure and properties as the phenomena to be explained.

In contrast, research where the network measures serve as the independent variable focus on the consequences of network structure. As defined by Hu et al. (2022), network effects are the desired impacts or consequences that a network's structure may have on the actors, the network itself, or the broader community the network serves. Thus, network effects concern the various decisions, outputs, and outcomes that may occur due to an actor's position in or the overall structure of the network. Such studies may ask, for example, whether an actor's network centrality is predictive of their success or if more densely connected networks exhibit higher performance outcomes (Provan & Milward, 1995; Raab, Mannak, & Cambré, 2015; Siciliano, Carr, & Hugg, 2021). Thus, network effects studies treat elements of the network as independent variables and question how these elements affect various outputs and outcomes of interest. Yet, despite growth in knowledge of network formation and effects, our field has lacked an action-oriented and practice-based approach to applying this knowledge. This article argues that we need to establish a third body of scholarship in public administration and policy on network interventions.

Network interventions provide a basis for the application of network research to practice. Network interventions are "purposeful efforts to use social networks or social network data to generate social influence, accelerate behavior change, improve performance, and/or achieve desirable outcomes among individuals, communities, organizations, or populations" (Valente, 2012, p. 49). The existing research on network interventions covers a vast array of areas and subject categories across the natural and social sciences but has primarily focused on the context of public health. Applications of network interventions have yet to reach public administration and policy, however.[1] We apply and extend the network intervention literature to develop an applied network science approach to public problem resolution. We argue the need to view networks as the context for intervention, where specific network mechanisms are identified for pragmatic control to better address governance and public management issues.

It is important to clarify here that network interventions are distinct from the use of networks as a form of organizing, such as via collaborative governance, to address complex policy challenges (e.g., Ansell & Gash, 2008). However, collaborative governance systems, when analyzed through a network lens as a set of nodes and relationships, could serve as a context for intervention. For example, if the trust-based ties among members in a collaborative governance regime were found to be very sparse, hindering their capacity for coordinated action, a network intervention could be deployed to increase the density of the trust network.[2]

[1] As a simple demonstration of the lack of uptake of network intervention in our field, we conducted a key word search in the title, abstract, and full text of all articles in ten prominent journals of public administration and policy. Only four articles were found to include the words "network intervention". The keyword search was conducted on April 20th, 2022. The journals searched were the following: Perspectives on Public Management and Governance; Journal of Public Administration Research and Theory; Public Administration Review; Public Administration; American Review of Public Administration; Policy Studies Journal; Journal of Policy Analysis and Management; Journal of Public Policy; Public Management Review; and Nonprofit and Voluntary Sector Quarterly.

Reconceptualization of a classic example clarifies the need for network intervention scholarship and practice in public administration and policy. One of the foundational studies on network effects conducted by Provan and Milward (1995) found that cities with more centralized mental-health delivery systems were better performing. Several other studies have supported this conclusion by linking centralization to performance (Raab et al., 2015; Sandström & Carlsson, 2008). If a city was seeking to improve its mental health service delivery and sought a more centralized network, how could this be achieved? What recommendations would we make? Translating Provan and Milward's (1995) insights into action requires identification and manipulation of network mechanisms to actively increase the service delivery network's centralization. To date, our field has not pursued such questions. Network interventions provide a means to do so. Grounded in a pragmatic, action-oriented approach (Mischen & Sinclair, 2009), network interventions link scholars with public managers, policymakers, and stakeholders in a democratic, participatory manner to identify network conditions and craft interventions to improve outcomes.

[2] As will be detailed in one example later in the article, it is also possible to think of the creation of a collaborative forum as a type of network intervention. For instance, consider a broader regional system of stakeholder relationships around a common policy issue or shared resource. If those relationships were mapped and found to be siloed, one potential intervention into the system could be the creation of a collaborative forum to facilitate the emergence of communication across the stakeholders. What makes network interventions different from other interventions or policy tools, is that they purposefully use network data and theory to understanding where, when, and how to intervene into existing systems.

This article aims to develop a typology to conceptualize strategies of intervention in public sector networks and to begin a scholarly discussion on the application of these strategies in our field. We aim to establish ideas on how public managers and leaders can facilitate the organization of networks in ways that are beneficial for addressing specific challenges and use existing network structures to spur changes in attitudes and behaviors. We rely on information and theory developed from existing research on network formation and network effects coupled with practitioner-based contextual insights to use networks for practical action. That knowledge provides the understanding needed to identify where and how to intervene in networks to improve their functioning, efficiency, and performance.

We begin by describing the core concepts in the theory of network interventions found primarily in the public health literature. Next, we expand the existing typology of interventions to include macro institutional strategies more suited to public administration settings. We then identify the extant public administration and policy literature that provides the building blocks for a framework for interventions. Finally, we classify the various modes of intervention along two dimensions and discuss how each strategy might be implemented in practice. As seen in the following section, the existing network intervention literature focuses on inter-personal networks. Thus, we apply these concepts to other types of networks common in public administration and policy. Such networks include inter-organizational networks (where organizations serve as the actors of interest) and governance settings where individuals often act as representatives of their organization.

**WHAT ARE NETWORK INTERVENTIONS?**

Network interventions are defined as the use of network theory and data to design interventions and achieve desired outcomes for a specific target population of actors. Network interventions have been categorized into four main strategies (Valente, 2012): *individual*, *segmentation*, *induction*, and *alteration*. Each strategy points to potential network processes and structures that may be manipulated to effect change regarding a specific problem. We summarize each of these strategies below.

The *individual* strategy is focused on the identification of champions or opinion leaders. The focal individual is identified based on knowledge of a network's structure and then relied upon to act as a change agent. These individuals promote and legitimize behavior based on their access and status in the network. While methods of identifying opinion leaders vary from rankings based on degree centrality to more sophisticated algorithms (Valente & Pumpuang, 2007), the goal is to detect the individual members most likely to prompt a behavioral or attitudinal change in others. For example, Starkey, Audrey, Holliday, Moore, and Campbell (2009) used a peer nomination questionnaire to identify individual students in schools to disseminate health messages to reduce the prevalence of youth smoking. The authors found that the network intervention strategy relying on change agents reduced the odds of being a regular smoker by 22% compared to control schools.

The *segmentation* strategy, rather than identifying particular actors, selects groups of actors to shape or change decisions at the same time. Such an approach is most suitable where behavioral change occurs at the group level, as with adopting a technology that would affect all group or sub-group members. Group-based learning and behavioral change strategies can be advantageous as group members can lend support, reinforce learning, and reduce the risk to individual adopters (Valente, 2010). As with the individual strategy, identification of the appropriate group typically relies upon the structure of the network. Group detection algorithms, such as Girvan-Newman (Girvan & Newman, 2002), can be applied to identify existing groups in the network. For example, Frank, Chen, Brown, Larsen, and Baule (2022) identified latent subgroups based on collegial ties in a natural resource management network. Once identified, they strategically convened meetings for those subgroups to focus on commonalities and plan future activities.

The *induction* strategy involves the use of existing social ties to diffuse beliefs or behavior from one actor to another. The induction strategy differs from those previously discussed as the emphasis shifts from a focus on the nodes (and their position in the network) to a focus on the edges. For example, Hoffman et al. (2013) conducted an HIV prevention intervention. They recruited injection drug users and randomly assigned some of them to receive training sessions on communicating risk reduction techniques to members of their network who also engaged in injection drug use. Relying on social influence and diffusion of information theory, Hoffman et al. (2013) tracked the HIV status of the network members and found evidence of reduction in incidence rates of those in the trained networks. Valente (2012) notes that word-of-mouth advertising and network outreach strategies are forms of induction as they rely on existing ties to diffuse ideas and information. Respondent-driven sampling methods, which rely on known individuals or clients using their network to recruit additional subjects, are also an induction strategy (Heckathorn, 1997). Respondent-driven sampling has proven to be an effective method for reaching hidden and hard-to-reach populations and has been deployed for several health promotion programs, including vaccinations (Valente, 2010; Valente et al., 2009).

Finally, the *alteration* strategy seeks to alter the existing structure and composition of the network. Given that a network consists of a set of nodes and the relationships among them, the alteration strategy has four primary approaches: node addition, node removal, link addition, and link removal. For example, an antiterrorism campaign might remove a critical node in order to degrade the terrorist organization's capability (Roberts & Everton, 2011). Similarly, researchers could identify links that bridge network holes. Removing such bridges would drastically decrease network cohesion and prevent the movement of disease or ideas from one part of the network to another. Or researchers could identify existing silos in a knowledge-sharing network and intervene to establish bridges to support greater knowledge flow (Frank et al., 2022).

Each of the four strategies use or alter the network structure to achieve some form of attitudinal, behavioral, or capacity change. The specific goals of any intervention, however, will vary dramatically. For example, one study may attempt to build greater connectivity among the actors in the network (e.g., an organizational intervention to break down silos), while another may aim to dissolve ties and fragment the network (e.g., a public health intervention to stop disease spread). Therefore, the strategies chosen depend on the relationship of interest, what is flowing or moving through the network, and the specific attitude or behavior sought to be changed (Valente, 2012). Thus, a particular network structure may be quite effective in some contexts and ineffective in others.

***Extending the Network Intervention Typology.*** As will be highlighted in the following section, there are several points of intervention in public administration and policy settings for which the current network intervention typology does not cover. First, most of the network intervention literature is focused on inter-personal networks (e.g., smoking cessation among school-aged children). In a later section, we develop several building blocks for thinking about the role of interventions in inter-organizational and governance contexts. Second, none of the existing strategies identified above suggest how interventions might stimulate or facilitate the evolution of a network, i.e. whole network dynamics (c.f. Provan, Fish, and Sydow 2007). Such an approach would be needed in situations, for example, where a city or a region sought to develop a more centralized network for public service delivery. While the alteration strategy provides micro-level interventions (individual node or link additions), it does not operate at the institutional or environmental level. For this, macro-level interventions are often needed that change the rules, penalties, or benefits of the actors in the network and thus may serve to change their tie formation decisions. For instance, one could alter the design of formal institutions, such as the compliance mechanisms or power-sharing arrangements to encourage collaboration among regional actors (Schlager, Bakkensen, Olivier, & Hanlon, 2021). One could also take a more direct approach by establishing formal rules about which organizations should work with one another (Olivier, 2019). We will discuss such macro-level strategies in more detail below.

To help organize and provide insight into the nature and range of intervention strategies, Figure 1 (adapted from Siciliano, Wang, Hu, Medina, and Krackhardt (2022)) connects existing research on network formation and effects to factors at the micro and macro level. An intervention strategy can be located in Figure 1 based on the level at which that strategy is enacted and operates (micro vs. macro) and on the direction of its intended effects, i.e., whether the intervention seeks to change aspects of network formation or if it seeks to use the existing network structure to affect outputs or outcomes. The figure places the network structure at the center, where the network is defined as the set of relevant actors and the ties that connect them.

At this level of abstraction, the intervention strategies are agnostic to the types of nodes or network.

On the network formation side, macro-level factors that can influence a network's structure include the policy domain, institutional rules, and aspects of organizational culture. Micro-level factors focus on the attributes of the actors that influence their tie formation decision. These attributes include attitudes, personality traits, resource-based capabilities, as well as interactions between attributes such as homophily. Incorporating sociological insights (e.g. Coleman's Boat (Coleman, 1990)), macro and micro features are connected, as institutional and environmental settings can shape the constraints and incentives actors face and thus shape their collaborative decisions.

On the network effects side, the overall structure and shape of the network can influence actor outputs and outcomes. For instance, more densely structured networks may allow information to diffuse faster and thus make individual actors more effective. Or actors in network bridging positions may be more likely to gain access to novel information leading to higher performance. At the same time, the overall structure of the network can have implications for its performance as a whole. While such macro effects are not the specific point at which a network intervention occurs, they are often the likely target or performance outcome driving interest in the use of interventions. Micro and macro outputs and outcomes are also connected on the network effects side of Figure 1, as patterns of micro-level behavior can have aggregate and emergent effects at the system level.

**Figure 1: Points of Intervention in Network Processes**

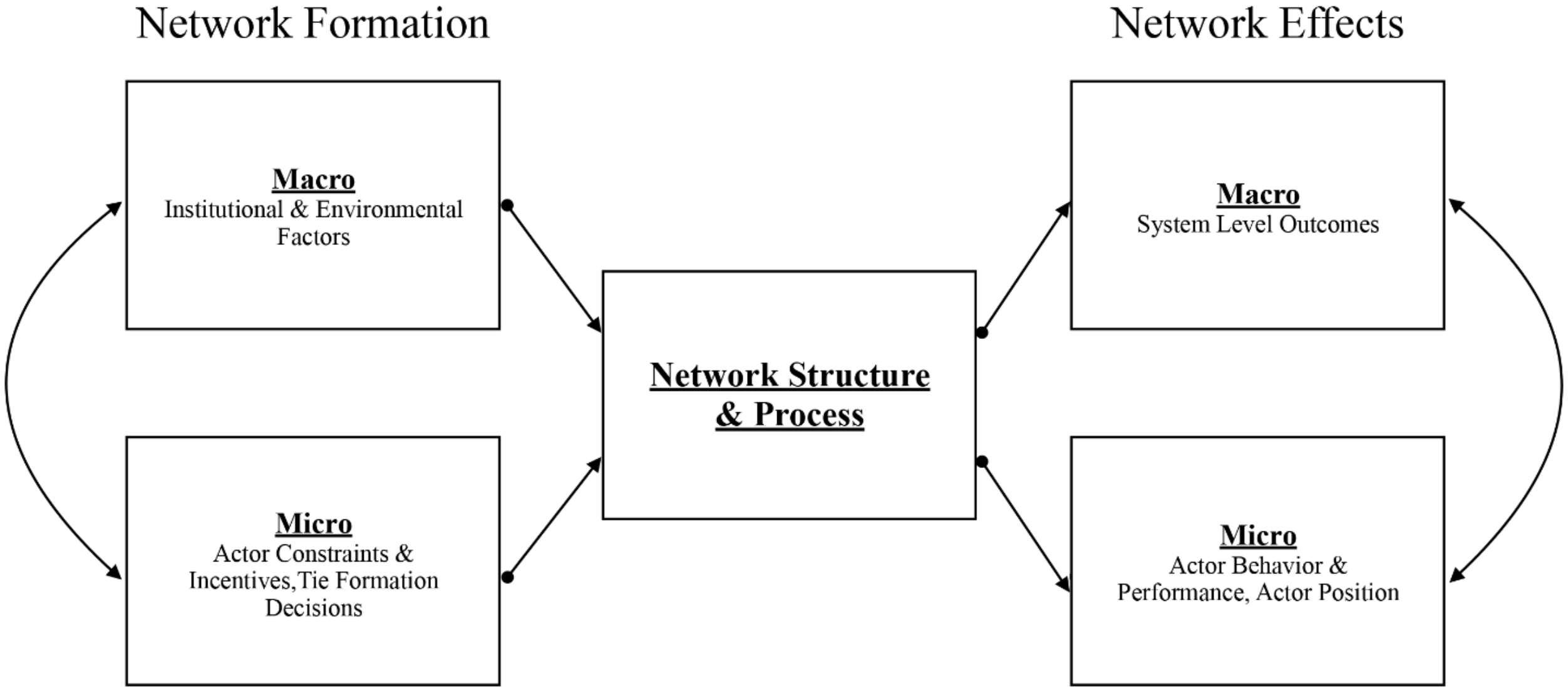

Based on Figure 1, we classify the individual, segmentation, and induction strategies as micro-level strategies on the network effects side. Scholars interested in group/cluster strategies might also conceptualize segmentation as a meso-level strategy. In regard to the individual and segmentation strategies, the intervention relies on an individual or set of actors strategically connected in the network. For these strategies, the network structure is assumed to be fixed, and optimal individuals or groups are then identified from that fixed structure. The induction strategy also assumes the network structure is fixed and relies on the existing ties of actors to diffuse beliefs and behaviors.

Although the alteration strategy also operates at the micro level, it is different than the other intervention strategies. When using an alteration strategy, the intervention is attempting to change the existing structure of the network by adding or deleting nodes and edges. Here, one is interested in modifying the network's structure in ways that may disrupt or advantage some policy relevant aspect of the network. While all interventions aim to influence individual and collective effects, we consider alteration strategies as focused on network formation because that is the point at which the intervention operates. Once the network shape or its membership has been changed, the results of that alteration (e.g., reduced cohesion) are expected to lead to intended effects (e.g., the slower spread of infectious disease). The critical point is that the action occurs on the network formation side for alteration strategies. Thus, rather than taking the network structure as given and using it to identify certain individuals or groups, this approach treats the network as dynamic and manipulable. The network changes established by the alteration strategy then lead to changes in individual and/or collective behavior.

All of the intervention strategies identified by Valente (2012) are micro-level strategies. We extend Valente's typology to include macro-level interventions. Macro-intervention approaches focus on incentives, opportunities, institutional norms, rules, and processes that influence the relationships among multi-sector organizations to achieve public policy outcomes (Ansell & Gash, 2008; Emerson & Nabatchi, 2015; Feiock, 2013; North, 1991; Ostrom, 2009). Indeed, there are many macro-level interventions designed to change behavior. To extend these studies into *network* interventions, the relevant question is: how do these interventions change the network members' tie formation strategies and thus the network's overall shape? We seek to identify the network mechanisms by which institutional interventions achieve public policy outcomes more explicitly. We label this type of intervention the *institutional* approach. Institutional intervention strategies may arise from the decisions of the actors in the network themselves or from a higher-level actor, such as a meta-governor (Gjaltema, Biesbroek, & Termeer, 2020). Table 1 summarizes the five strategies and the level (micro or macro) and direction (network formation or network effects) at which that strategy operates. In the following section, we will establish a set of building blocks to support the development of strategies and mechanisms for intervention in public sector settings.

**Table 1: Summary of Intervention Strategies**

| Intervention Strategy | Level | Direction | Definition |
|---|---|---|---|
| *Individual* | Micro | Network Effects | Identify actors in structurally optimal positions to serve as opinion leaders and change agents. |
| *Segmentation* | Micro/Meso | Network Effects | Identify groups of actors to shape or change decisions at the same time. |
| *Induction* | Micro | Network Effects | Activate a process on the network or engage in network outreach. |
| *Alteration* | Micro | Network Formation | Alter the structure of the network through the addition and removal of nodes and/or links. |
| *Institutional* | Macro | Network Formation | Change the rules, benefits, and constraints faced by network members to influence tie formation decisions. |

*Implementing Network Interventions*. While network interventions are designed based on theories from psychology, sociology, and network science, the ultimate authority to enact network interventions in real world settings resides with policymakers and public managers. Existing examples of network interventions in public health tend to pair university researchers/social network analysts with health organizations and practitioners. Guided by a practice-based problem (Carboni et al., 2019), such as improving mental health service delivery, the network analyst would map the relevant network(s). Then, relying on practitioner knowledge coupled with concepts and theories from network formation and network effects research, the analyst would work with the community/practitioners to design an intervention tailored to the observed situation and network structure. We envision a participatory orientation where the researcher plays an intentional role in the design, implementation, and evaluation of the network intervention, as opposed to a technocratic role that divorces theory from practice (Mischen & Sinclair, 2009). Thus, we envision the role of the network analyst as working closely with practitioners and stakeholders in a participatory fashion to analyze and address public problems (Ansell, 2011; Shields, 2008; Whetsell, 2013). However, this approach also requires attention to the ethical consequences of interventions, especially when vulnerable policy targets are at issue.

## BUILDING BLOCKS FOR PUBLIC SECTOR NETWORK INTERVENTIONS

Existing applications of network interventions focus on individual people as the nodes of interest. This includes work on behavior and attitudinal change which take advantage of the important role social networks play in shaping human behavior. Building on this work, this section considers literature relevant for intervention into governance and inter-organizational settings. Such settings include situations where organizations, as well as individuals acting as representatives of their organizations, serve as the node of interest and work together to address complex policy and management issues. Due to space constraints, we explore four areas of scholarship relevant to network interventions in the public sector: (i) network governance and governance networks; (ii) institutional analysis and development; (iii) new institutionalism and strategic action fields; (iv) collaborative governance regimes. Our intent in exploring these select few literatures is to demonstrate the insights and ideas these existing frameworks and theories offer for network interventions in public networks. Scholarship based on these frameworks has been used to support and justify various mechanisms associated with network formation (Siciliano, Wang, et al., 2021) and network effects (Hu et al., 2022). Our intent is not to establish a competing framework of network governance or collaborative management.

Rather, we seek to use these existing frameworks and theories to identify potential opportunities to promote behavior change and improve outcomes in management and policy settings.

Following Rhodes (2007), *governance* is defined as steering self-organizing, cross-sector, purpose and resource-oriented inter-organizational relationships. This definition fits a standard definition of inter-organizational networks. Here we borrow Provan and Kenis' (2008) definition of *inter-organizational networks*: "groups of three or more legally autonomous organizations that work together to achieve not only their own goals but also a collective goal. Such networks may be self-initiated, by network members themselves, or mandated or contracted, as is often the case in the public sector" (p. 231). The organizations constitute the network actors (nodes), and the relationships between the organizations constitute the network connections (edges). Public administration and policy relevant networks tend to involve actors from different sectors, of different sizes, with heterogenous resource endowments, where relationships take different forms, such as formal partnerships (PPPs), contracts, agreements, or even simple information exchange relations.

Jones, Hesterly, and Borgatti (1997) explain why network forms of governance tend to emerge. They integrate two previously disparate bodies of theory: transaction cost economics and social network theory. When transactions between organizations have high frequency, uncertainty, asset specificity, and task complexity, organizations will tend to engage in contracting instead of spot exchanges on an open market. Long-term contracting gives rise to "relational embeddedness" within the organizational dyad. As dyads aggregate, relational embeddedness evolves into "structural embeddedness" where numerous organizations are now directly and indirectly connected within a broader inter-organizational network. At this point, social mechanisms such as reputation and collective sanctions emerge to adapt, coordinate, and safeguard the network itself. The regulation of these high-level social mechanisms within a structurally embedded context is called “network governance.” The governance tasks of regulating these high-level structures constitute the unique activities of steering the emergent properties of a network at the macro-scale.

In the public sector, organizations do not cooperate primarily because they wish to economize on transaction costs or to jointly produce a good or service to realize competitive advantage, but rather to resolve some broader public problem. Public sector organizations wield greater legal authority but grapple with a broader, more subjective set of problems. Provan and Kenis (2008) advance theory regarding the governance of inter-organizational networks for network effectiveness. Reviewing numerous cases in public sector network governance, they apply the concepts of brokerage and participation to identify three forms of governance: participant-governed, lead organization-governed, and governance by network administrative organization (NAO). From here, they suggest each governance form should be aligned with a context to achieve network effectiveness.

The above studies provide useful concepts regarding network governance, and offer insight on possible interventions into governance systems. Such interventions could consist of changes in transaction costs, reductions in uncertainty, or the introduction of a network administrative organization. Even though public sector authority over inter-organizational networks is considerably greater in public versus private contexts, there are still important limitations to the authority of policymakers and public managers, and hence limits to the role of agency in such networks. These limitations point toward the more general “paradox of embedded agency,” which characterizes the classic agency-structure debate regarding the limits of actors embedded within institutional fields to achieve change (DiMaggio & Powell, 1983; Fligstein, 2001; Garud, Hardy, & Maguire, 2007; Hardy & Maguire, 2017). Given the limitations of agency and authority, the network governance literature often does not provide prescriptive suggestions about network interventions. Moreover, the lack of authority within cross-sector networks means policymakers and administrators are often limited to leveraging macro-level institutional mechanisms. Thus, demonstrating the need to extend Valente's (2012) typology to include an institutional intervention strategy.

North (1991) defines institutions as formal or informal constraints that structure social exchange, such as laws, rules, and social norms. Ostrom's (2009) work showed that institutional level rulemaking, such as establishing boundaries, property rights, enforcing sanctions, and setting quotas, when tailored to the local context, may be appropriate governance mechanisms for reducing barriers to self-organization. Ostrom's work focuses more broadly on developing institutions rather than direct intervention in the mechanisms of a network. Building on this work, the institutional collective action framework identifies a number of mechanisms for "integrating" collective action dilemmas, such as contracts, partnerships, mandated agreements, but also informal networks, constructed networks, and multiplex self-organizing systems (Kim et al., 2021). In each of the mechanisms identified in the institutional collective action framework there exist a set of relations among the actors (a social or inter-organizational network). Thus, whether mandated agreements, contracts, or partnerships are chosen as the mechanism to solve a collective action dilemma, the social and professional relations that tie the actors together remain and influence their individual and collective success. Therefore, network interventions may be leveraged by policymakers or public managers to achieve specific ends in a variety of governance settings.

Delving further into institution-oriented theories, "new institutionalism" and the theory of strategic action fields have recently been invoked to characterize policy implementation in inter-organizational settings (Moulton & Sandfort, 2017; Sandfort & Moulton, 2020). According to DiMaggio and Powell (1983), organizational fields characterize the population of potentially interacting organizations, placing environmental constraints upon them and homogenizing them through coercive, mimetic, or normative isomorphism. Fligstein (2001) and Fligstein and McAdam (2011) extend these insights by emphasizing the agency of the individual actor in pushing upward against field structure. As the theory goes, the actor is often left out of the picture but possesses strategic capabilities (skills) for motivating other actors toward collective strategic action. The strategic component refers to the goal-directed behavior of actors, as they form cooperative inter-organizational clusters to compete with other groups of actors.

Overall, while the institutional literature provides important insight into how rules and norms shape collective behavior, little work has explicitly explored the effects of institutional interventions on network structure. As previously noted, network intervention strategies are often quite direct, such as identifying network actors to disseminate information or resources, targeting specific nodes or edges for support, or creating or eliminating nodes or linkages in a network. This is in contrast to the focus of institution-oriented frameworks, which primarily focus on broader concepts like autonomy, authority, beliefs, and transaction costs (Feiock & Scholz, 2010; Kim et al., 2021). As Scott and Ulibarri (2019) point out, public administration studies focusing on institutional elements but lacking explicit network analysis can be categorized as "implicit network analysis."

Collaborative governance, as separate from the network governance literature, also provides an important source of concepts. Both collaborative and network governance share an interest in the management of institutional-level mechanisms that affect public problems (Wang & Ran, 2022), but collaborative governance is more focused on the elements of the decision-making process among a collection of stakeholders, e.g. power asymmetry, inclusivity, face-to-face communication, and trust (Ansell & Gash, 2008). With few exceptions (Scott, Thomas, & Magallanes, 2018; Ulibarri & Scott, 2016), the collaborative governance literature features little emphasis on the explicit elements of a network in which collaborators are structurally embedded (Wang & Ran, 2022). Collaborative governance tends to be focused implicitly on the structure, function, and dynamics of a network (Scott and Ulibarri, 2019). Emerson, Nabatchi, and Balogh (2012) provide an integrated framework of collaborative governance regimes, where "network connectedness" is mentioned as part of the system context.

However, rather than being one element of the system context, we argue the network itself is the context upon which other collaboration components such as drivers, motivations, and capacities, are enacted. Changing the rules, norms, or strategies of principled engagement will have implications for the structure of the relationships among the actors. Taking a more explicit approach, leaders of a collaborative governance regime could map the communication or trust relationships among the members of the regime. If there are isolated subgroups, say government actors from community groups, network interventions could be deployed to improve connectivity across the groups.

Table 2 captures how the theory and frameworks briefly discussed here might feed into a network intervention design. The table summarizes the information in this section, provides a brief description of each framework, and then assesses the network intervention content. Each framework contributes important elements that may be used as building blocks for the design and implementation of network interventions.

Table 2. Application of Extant Theory to Network Intervention

| Theoretical Framework | Description | Network Intervention Content |
|---|---|---|
| Network Governance & Governance Networks | Steering of multi-sector inter-organizational networks. Attention to fit between governance modes and network structures and processes. | Explicit mapping of networks. Emphasis on network properties, dynamics, and management. Network treated as context. Explicit analysis of network formation and effects. Lacks explicit design and implementation of network interventions. |
| Institutional Analysis and Development & Institutional Collective Action | Structuration of high-level factors, rules, norms, rights, etc. to facilitate self-organization among multi-sector inter-organizational relations. | Occasional mapping of networks. Mostly implicit or metaphorical attention to network formation and effects. Network treated as one of several tools/choices for resolving institutional collective action dilemmas. Lacks explicit design and implementation of network interventions. |
| New Institutionalism & Strategic Action Fields | Focus on agency, skills, culture, and structure in motivating cooperation between actors. Inter-organizational relations are conceived as meso-level fields. | No explicit mapping of networks. Network treated as context (fields). Mostly implicit or metaphorical attention to network formation and effects. Provides insight into the context of intervention. Lacks explicit design and implementation of network interventions. |
| Collaborative Governance Regimes | Focus on inputs and processes of interaction (inclusion, trust) in a cooperative multi-sector inter-organizational setting. | Limited mapping of networks. Network mostly treated as a form of organization bringing stakeholders from multiple sectors and jurisdictions together. Emphasis on information sharing, consensus building, and design. Mostly implicit or metaphorical attention to network formation and effects. Lacks explicit design and implementation network interventions. |

## A TYPOLOGY FOR PUBLIC SECTOR NETWORK INTERVENTIONS

Thus far, we have drawn attention to two sets of factors that frame our thinking on the application and development of network interventions in public administration: 1) the difference between macro-institutional mechanisms, and micro-network mechanisms, and 2) the choice of emphasis on network formation versus network effects. These two dimensions are arrayed below in Table 3. The columns depict whether the intervention is focused on network formation versus network effects. The rows distinguish between the macro-level of network structure which refers to whole network/group properties and dynamics, and the micro-level, which refers to the properties and dynamics of the individual actors and relationships within the network. Within the cells, the different macro and micro-level strategies are identified by the particular intervention type along with sample tactics suitable for public administration and policy.

**Network Interventions Targeting Network Formation.** When researchers or practitioners seek suitable strategies for intervening in networks, one of the first decision points is whether the intervention is aimed at network formation or network effects. As An (2011) argues, actors can design strategies to alter the network structure or they can utilize the network structure to facilitate attitudinal or behavioral change. If the current structure of the network is deemed suboptimal (i.e., the shape or composition is ill-suited to the current environment), then one would look to strategies on the network formation side of Table 3. Here, two intervention categories treat the network as dynamic. One can take a micro *alteration* approach to rewire specific ties in the network or add/remove certain nodes. Or one can change the *institutional* environment, where no specific node or tie is targeted for addition or removal, but rather incentives or constraints are implemented to alter dyadic behavior leading to changes in the overall structure of the network.

**Table 3: Typology for Public Sector Network Interventions**

| Level of Network | Network Formation | Network Effects |
|---|---|---|
| **Macro-Level Structure** | *Institutional*<br>Establish or adjust norms, rules, regulations, compliance mechanisms, power-sharing arrangements, taxes, sanctions, property rights, boundaries, and quotas. Create new institutions such as collaborative forums and platforms.<br><br>**Example**: decrease legal barriers to increase cooperation between organizations on the production of strategic resources. | |
| **Micro-Level Structure** | *Alteration*<br>Removal of or requirements for particular organizations to participate in mandated networks; adoption of NAOs; establishment of contracts, alliances, PPPs, consortiums, and interlocal agreements.<br><br>**Example**: regional governments establish a network administrative organization to coordinate numerous public-private partnerships for large scale infrastructure projects. | *Individual*<br>Target specific organizations for adoption of behavior, diffusion of innovation, information, or resources.<br><br>**Example**: identify a well-respected government agency, non-profit, or private organization to diffuse the adoption of energy efficiency programs by local government.<br><br>*Segmentation*<br>Target groups of organizations for behavioral change, diffusion of innovation, information, or resources.<br><br>**Example**: target an industry consortium to adopt standards and benchmarks.<br><br>*Induction*<br>Initiate word of mouth information diffusion and viral campaigns.<br><br>**Example**: local government use of social media to promote factual COVID-19 information spread. |

Examples of interventions at the macro-institutional level, include network-wide rules, norms, sanctions, property rights, geographical boundaries, and quotas. For example, changes in anti-trust laws made by the 1984 National Cooperative Research Act are widely recognized as contributing to substantial increases in inter-organizational cooperation in the high-technology sector (Jorde & Teece, 1990). These changes could be reconceptualized in network intervention terms as an institutional intervention that directly affected the structural cohesion of the network. Examples of a micro-alteration level intervention might include establishing network administrative organizations, or establishing or supporting public-private partnerships, or terminating partnerships. For instance, Whetsell, Siciliano, Witkowski, and Leiblein (2020) examine the case of Sematech, a U.S. Department of Defense sponsored research consortium active in the late 1980s-1990s. They reconceptualize Sematech as a network administrative organization (NAO) and argue that NAOs can catalyze self-organization in cooperative networks through an alteration (i.e., link addition) strategy.

To better illustrate the process of network intervention on the network formation side, we offer a simulated example in a governance context. Consider an institutional intervention among stakeholders within a watershed. The watershed encompasses a large number of communities, recreational parks, industrial areas, and farms. To explore growing water quality issues, a concerned local government worked with a researcher to assess the policy landscape and understand the existing patterns of trust and communication among the different stakeholder groups. Figure 2a depicts the pre-intervention network map of communication relations among the key stakeholders in the watershed. The nodes are shaded based on membership in one of the four primary stakeholder groups: community members, business/industry representatives, farmers, and local governments. The network in Figure 2a reveals a strong tendency to form ties only with others within one's stakeholder group, with very few connections across groups. Networks shaped in this manner foster loyalty and identity to one's subgroup. Consequently, when searching for solutions to problems confronting the broader watershed, actors in networks structured like the one in Figure 2a tend to focus on solutions beneficial to their group and not the collective, hurting overall capacity and performance (Krackhardt & Stern, 1988).

**Figure 2: Network Diagrams of Pre and Post Institutional Intervention.**

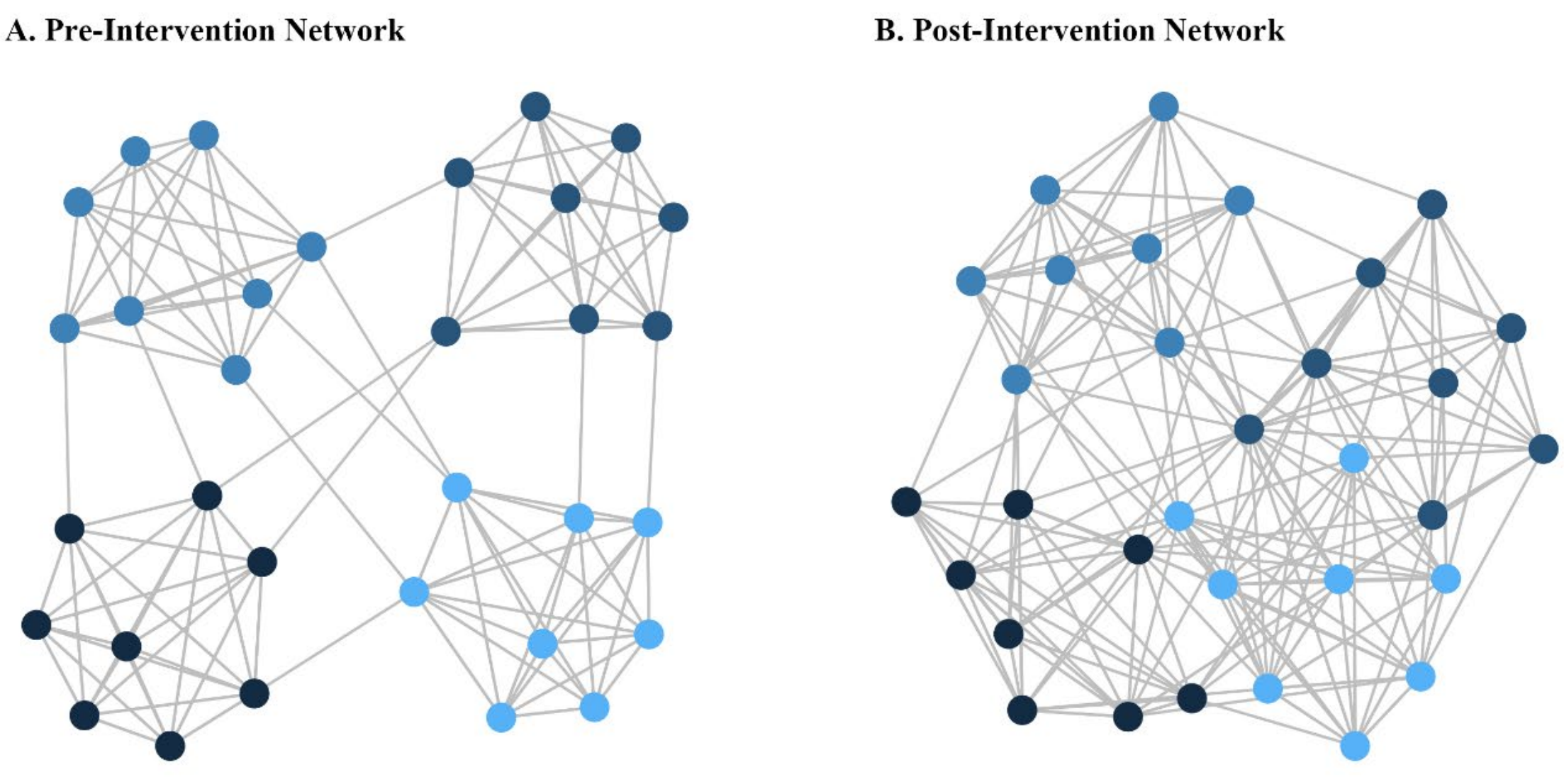

**Figure Notes:** *The network in Figure 2.a demonstrates the existing patterns of relationships among four key stakeholder groups associated with a watershed. Figure 2.b reveals the patterns of interaction after the establishment of an institutional intervention in the form of a collaborative forum. The intervention was designed to improve the interaction and communication among the stakeholder groups, moving from strongly homophilic to more cross-group connections.*

Based on this knowledge of the existing network, the concerned local government decided to establish a collaborative forum to provide a space and opportunity for the stakeholders to have principled engagement through face-to-face dialogue and trust-building activities (Ansell & Gash, 2008; Emerson & Nabatchi, 2015; Scott, 2016). Before establishing this collaborative forum, the stakeholders had limited opportunity for meaningful dialogue and interaction with one another, leading to little chance of system-wide coordination and collaboration. The establishment of the collaborative forum changed existing rules and norms of interaction to focus on consensus-based decision-making and collective benefits. After several months of regular meetings hosted by the collaborative forum, the communication relations among the stakeholders were measured again. Figure 2b represents the new set of ties. Though there remains a tendency to interact within one's own group, there is also a large increase in the number of cross-group relations. While facilitating communication and building trust are essential activities of public managers, what is important to remember here is that both the impetus for the intervention and the assessment of its effect are based on knowledge of the network's structure. The intervention was based on insight into how institutional changes within an existing system might alter the existing network structure.

**Network Interventions Targeting Network Effects**. Three different approaches have been identified on the network effects side: individual, segmentation, and induction. The individual and segmentation approaches use knowledge of the network's current structure to identify specific actors or groups to target. For example, the *individual* intervention identifies key influencers to diffuse desired information and behavior to other actors.

In public administration and policy, public managers often wish to diffuse a particular behavior among employees or among other organizations within a cooperative network. However, public managers operating within inter-organizational networks often lack the formal authority to force others toward the desired action. The *individual* strategy suggests that a key actor may be an effective approach for promoting the behavior within the network. Choosing the actor, however, is the critical part. While the most supportive actor of a reform may be an obvious choice to lead implementation, their influence could negatively affect implementation if that actor is not well connected in the informal network (Krackhardt, 1992). Research suggests the value and need for explicit network analysis and network-based interventions to effectively promote policy change and adoption (Leonardi & Contractor, 2018).

Consider Figure 3a, depicting a hypothetical communication network among city managers in a large metropolitan region. Within this region, assume that the COG (council of governments) is promoting a sustainability initiative to reduce greenhouse gas emissions. This initiative is in the form of a compact that establishes goals on climate action that can be taken at the municipal level to promote sustainability and climate adaptation in the broader region. The Director of Sustainability of the COG would like each municipality in the region to sign onto the compact and agree to meet the climate goals set forth. The Director of Sustainability believes that having more signatories to the compact would significantly improve the efficiency and equity of the initiative. Yet, the COG lacks formal authority to compel municipalities to sign. Moreover, given the changes required to meet the goals of the compact, there is resistance among many municipalities. Some, however, are supportive.

Within this setting, assume the COG partners with a local government researcher who studies municipal collaboration. Like many studies conducted in public administration, the researcher could conduct a network analysis of the communication and existing collaborative ties among the city managers in the region. That study could also gather information on the manager's current support for the compact (potentially as a reflection of resident/citizen support). Such data could be used to produce the visualizations in Figure 3. The nodes in Figure 3a are scaled based on the city manager's support for the compact; the larger the node the higher the support. Seeking a champion or opinion leader for the compact, an obvious choice for the COG would be to select node A, highlighted by the arrow, who is known by their Director of Sustainability to be the most supportive of the compact. However, this city manager is not well-connected in the region. From a network intervention perspective, an optimal manager to lead the effort would be node B, again highlighted by an arrow in Figure 3b. In Figure 3b, the nodes have been rescaled based on in-degree to visualize the most central actors. Node B has a relatively positive assessment of the compact and is the most highly connected manager in the network. The numerous direct and indirect ties held by node B serve as conduits to promote the value of the compact by a trusted colleague. Krackhardt (1992) provides an instructive empirical example of this strategy by looking at a failed unionization attempt and the selection of a reform champion.

**Figure 3: Network Diagrams of an Individual Intervention.**

**A. Size Scaled on Node Support** **B. Size Scaled on Node Centrality**

Figure Notes: *Figure 3 plots the relationships among city managers in a region where the council of governments (COG) is looking to increase the number of signatories to a climate change compact. One strategy would be to select the most supportive of the compact, which is node A in Figure 3.a. An alternative strategy, based on network interventions, would be to select the most highly connected and trusted actor in the network, node B, as seen in Figure 3.b.*

Depending on the structure and composition of the network, an alternative individual strategy could be to identify popular and centrally located actors with negative impressions of the compact. Thus, rather than identifying a champion and using the champion's network popularity to spread positive influence, this approach identifies nodes that may be potential roadblocks in the uptake of reforms and targets them with additional materials and resources in hopes of changing their opinion.

Staying with the network effects side, the *segmentation* approach works similarly to the individual approach, but rather than identify individual actors as either champions or roadblocks, the segmentation approach identifies groups of actors to change at the same time (Valente, 2012). Groups in networks can be identified through visual inspection of the network map or clustering algorithms available in many social network analysis packages. In keeping with the sustainability initiative example, targeted groups could be those at the core of a core-periphery network. By providing core actors with additional resources and information on the reform, this strategy can allow their influence, via their high connectivity, to spread to those on the periphery. Alternatively, perhaps the network is separated into two components with little connection between the two. If one component has low perceptions of the initiative and is hesitant to adopt it, the entire group comprising that component could be targeted. Targeting groups for learning and behavioral change can be advantageous as peers can support and reinforce learning while also reducing the perceived risk of adoption (Valente, 2010).

## DISCUSSION

In this article we have advanced network interventions as a specific activity for implementation of programs and policy in the interconnected context of network governance.

Effective implementation of policy in the context of inter-organizational governance entails discrete choice between alternative modes of network intervention that are selected through analysis of network structure, social processes, and expected outcomes. We advanced a general typology categorizing intervention strategies based on 1) the processes targeted for change, network formation versus network effects; and 2) the level at which the intervention operates, micro versus macro. This categorization provides public administration and policy scholars and practitioners a useful way to understand where, why, and how a particular intervention might be deployed. Figure 4 summarizes the relationships between network formation, network effects, and network interventions.

**Figure 4: Linking Network Formation, Effects, Structure, and Interventions**

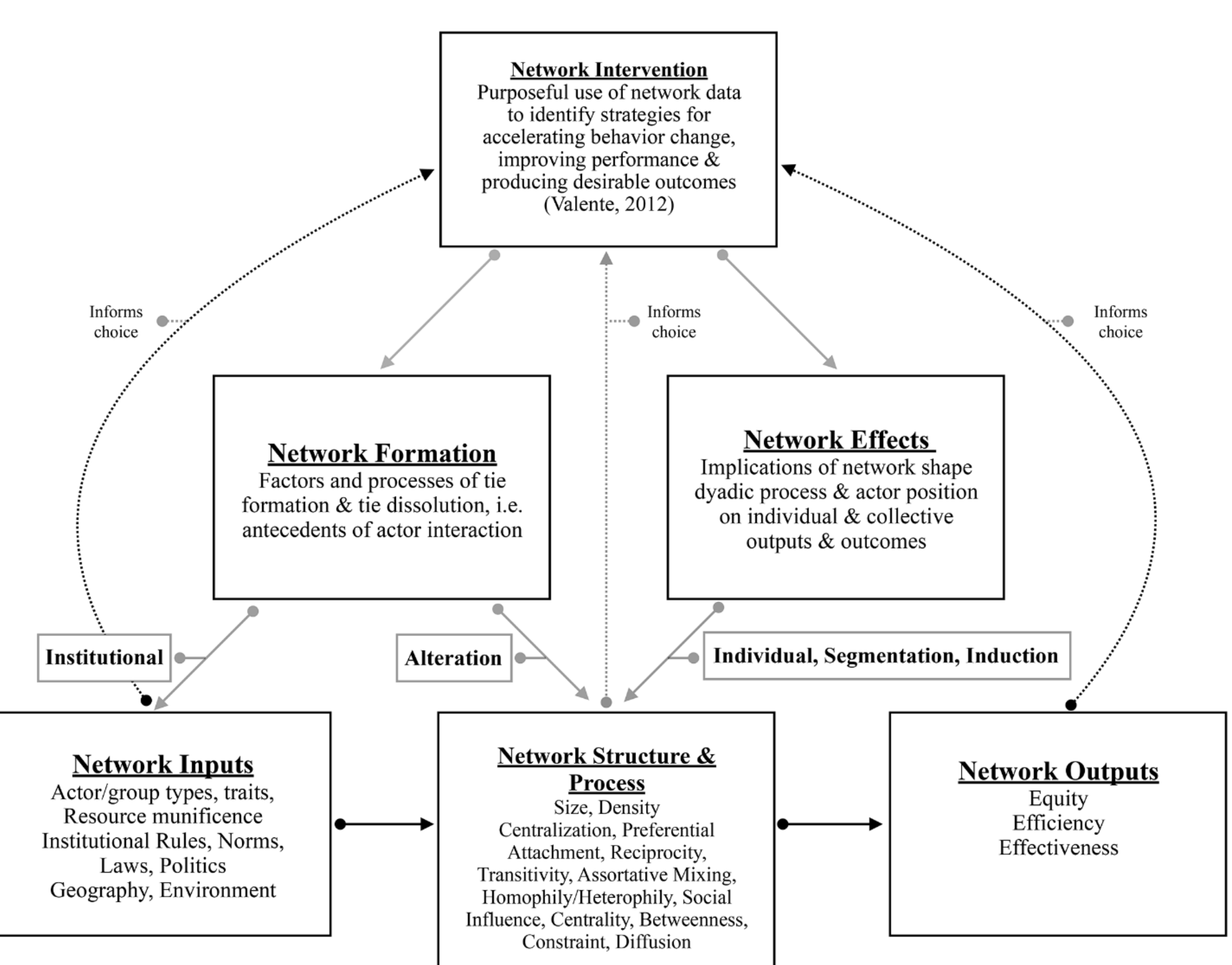

At the bottom of Figure 4 is a basic logic model for networks consisting of network inputs, network structure and processes, and network outputs. The logic model features of networks are connected to the two primary areas of scholarship: network formation and network effects. Network formation connects network inputs to the structure and processes of the networks that result through actor interactions. Network effects link those network properties to outputs. To date, research in public administration and policy has focused on describing relations or testing inferential hypotheses that connect inputs, to structure, to output. However, very little research has attempted to intervene into the system.

A critical next step is to establish a third area of scholarship on *network interventions*. This third area of scholarship is needed to move us beyond descriptions and inferences of network processes to actual interventions in these systems. As Lubell (2013, p. 547) argued, if our governance systems are "immune to purposeful intervention, the policy sciences will largely be confined to description and causal analysis rather than applied usefulness." Our goal in this paper is the pursuit of usefulness. We aim to spur research to use existing knowledge and frameworks to understand where and how interventions might occur in public sector networks to create public value.

Despite the promise of usefulness for network interventions, several challenges should be noted when considering their design and implementation. Network scholars in public administration and policy (working in conjunction with public managers and policymakers) must have a clear theory and understanding of the mechanisms at work in the network to intervene successfully. Thus, a clear articulation of the specific mechanism operating in a given context is necessary to align the network intervention with the relevant dynamics driving network formation and behavioral responses. Another challenge is in choosing the correct intervention strategy. Valente (2017) argues that future research needs to compare intervention strategies to identify the optimal approach in a given setting. Network interventions also require careful research designs. Whereas public administration tends to study single networks, evaluating the success of an intervention often requires control groups and experimental or quasi-experimental designs.

It is also important to note that there are limitations to what can be achieved in inter-organizational and governance networks. Here, limits to authority mean that policymakers and public managers often cannot, for example, simply add or remove organizations from a multi-sector network. In addition, public organizations have legal mandates and sovereign responsibilities to particular jurisdictions. As a result, they usually cannot simply decide to remove themselves from a network or insert themselves in other areas of a network with conflicting jurisdictions. Further, as the network governance, institution-oriented, and collaborative governance literature reviewed above has shown, the lack of traditional hierarchical controls in inter-organizational settings leads to complex power dynamics that can be challenging to steer toward public policy aims. Despite these challenges, this article has sought to demonstrate the opportunities for network intervention that exist in inter-organizational and governance networks.

## CONCLUSIONS

Network interventions offer a means to capitalize on the strengths of our existing theoretical and empirical work while pushing the field forward to consider networks as the point of policy and management intervention. An intervention approach to networks also helps to demonstrate how the decisions and policies made at both the micro and macro level can have important implications on social structure and how knowledge of that social structure can be used to enact and expedite change. Often managers and policymakers implement change without consideration of the specific network structures through which that change will occur. This treats the network as a black box and ignores the structural implications of those decisions. As argued in this article, policy changes are enacted on, operate through, and adjust the social networks that serve as the backbone of network governance. Our hope is that this article serves as a starting point for scholarship and implementation of network interventions in public administration and policy. Network interventions hold promise for our field's ability to purposefully use networks to improve social, organizational, and community outcomes through a range of micro and macro-level strategies.